\documentclass[preprint2]{aastex}

\slugcomment{Draft 1}
\begin{document}

\def\arcdeg{\hbox{$^\circ$}}
\def\arcmin{\hbox{$^\prime$}}
\def\arcsec{\hbox{$^{\prime\prime}$} }
\def\s8{\hbox{S$_{850{\mu}m}$ \ }}
\def\h{\hbox{$^h$ \ }}
\def\ltsim{\mathrel{\hbox{\rlap{\hbox{\lower4pt\hbox{$\sim$}}}\hbox{$<$}}}}
\def\gtsim{\mathrel{\hbox{\rlap{\hbox{\lower4pt\hbox{$\sim$}}}\hbox{$>$}}}}

\title{The Canada-UK Deep Submillimeter Survey VI: The 3-Hour Field.}

\author{T.M. Webb\altaffilmark{1}, S.A. Eales\altaffilmark{2}, S.J. Lilly\altaffilmark{3}, D.L. 
Clements\altaffilmark{2}, L. Dunne\altaffilmark{2}, W.K. Gear\altaffilmark{2}, H. Flores\altaffilmark{4}, M. Yun\altaffilmark{5}}

\altaffiltext{1}{Department of Astronomy and Astrophysics, University of Toronto, 60 St George St, Toronto, Ontario, M5S 3H8, Canada}
\altaffiltext{2}{Department of Physics and Astronomy, Cardiff University, P.O. Box 913, Cardiff, CF2 3YB, UK} 
\altaffiltext{3}{Herzberg Institute for Astrophysics, Dominion Astronomical Observatory, National Research Council}
\altaffiltext{4}{Observatoire de Paris, Section de Meudon, DAEC, 92195 Meudon Principal Cedex, France}
\altaffiltext{5}{Department of Astronomy, University of Massachusetts, Amherst, MA 01003, USA}
\begin{abstract}
We present the complete submillimeter data for the Canada-UK Deep Submillimeter Survey (CUDSS) 3\h field.  The observations were taken with 
the Submillimeter Common-User Bolometer Array (SCUBA) on the James Clerk Maxwell Telescope (JCMT) on Mauna Kea.     The 3\h field is  one 
of two main fields in our  survey and covers 60 square arc-minutes to a 3$\sigma$ depth of $\sim$ 3 mJy.  In this field we have detected 27 
sources above 3$\sigma$, and 15 above 3.5$\sigma$.  We assume the source counts follow the form $N(S) {\propto} S^{-\alpha}$ and measure 
$\alpha$ = 3.3$^{+1.4}_{-1.0}$. This is in good agreement with previous studies and further supports 
our claim  \citep{eal00} that SCUBA sources brighter than 3 mJy produce $\sim$20\% of the 850$\mu$m background energy.   Using preliminary 
ISO 15 $\mu$m maps and VLA 1.4 GHz data we have identified counterparts for six objects and have marginal detections at 450$\mu$m for two
additional sources.     With this information we estimate a median redshift for the sample of 2.0$\pm$0.5, with $\sim$10\% lying at $z<$ 1.
We have measured the angular clustering of  \s8$>$ 3 mJy sources using the source catalogues from the CUDSS two main fields, the 3\h and 14\h 
fields, and find a marginal 
detection of 
clustering, primarily from the 14\h field, of $\omega(\theta)=4.4\pm2.9  \theta^{-0.8}$.  This is consistent with clustering at least as strong as 
that seen for the Lyman-break galaxy population and the Extremely Red Objects.  Since SCUBA sources are selected over a broader range in 
redshifts 
than these two populations the strength of the true spatial clustering is expected to be correspondingly stronger.

\end{abstract}


\section{Introduction}

Over the last decade there have been great steps forward in our understanding of the formation and early evolution of galaxies.  There are 
currently  two  general, distinct theories of massive galaxy formation, though the true picture is likely a combination of the two.   In the first, 
galaxies form  over a range of redshift, from the gradual hierarchical merging of smaller aggregates \citep{bau96,kau98}. In this picture galaxy 
formation is an ongoing  process characterised  by star formation rates of moderate magnitude. In the second scenario galaxies form at 
high-redshift on short timescales from the collapse of a single  object and undergo one massive burst of star formation \citep{egg62}.  They then 
evolve passively to galaxies of the  present-day. 

The observational picture is still  somewhat confused. Optical studies have found that the  luminosity density of the universe increases  
out to z$\sim$1 \citep{lil96,mad98,hog98} and, according to the observations of the   Lyman-break galaxy population (LBG), does not decrease to 
at least   z$\sim$4 \citep{ste99}. However,  the star formation rates in individual LBGs of a few tens of solar masses per year \citep{ste96}, 
though large compared to local starbursts, are too moderate  to form an elliptical galaxy on  a dynamical timescale of $\sim$10$^8$ 
years, and suggest gradual, hierarchical  formation. The hierarchical model  is further supported by the  increase in the  rate of galaxy-galaxy 
interactions 
with 
redshift \citep{pat01}.

On the other hand, the spheroids, which contain $\sim$2/3 of the stars in the universe \citep{fuk98}  still appear to be old at z$\sim$1 
\citep{zep97,cim99,sco00,mor00}.  The homogeneous nature of their  stellar populations today \citep{bow92} imply formation over a  short 
timescale and at high redshift. However, until recently, no high-redshift object with star formation rates large enough to form a 
massive spheroid in a dynamical timescale had been seen.  The deep submillimeter surveys of the last five years, \citep{sma97,hug98,bar98,eal99} 
with the Submillimeter Common-User Bolometer Array (SCUBA) on the James Clerk Maxwell Telescope (JCMT), have uncovered just such a 
population.  \

 The results of these deep SCUBA surveys have been exciting and in general agreement with each other.  The population revealed by the SCUBA 
surveys  
covers a broad range of redshift, with a median  redshift of 2 $<$ z $<$ 3 \citep{eal00,dunl01}.   Many of the  submillimeter detections which have 
secure 
optical/near-infrared  (NIR) counterparts show disturbed morphology or multiple-components suggestive of galaxy mergers \citep{lil99,ivi00}.  
These 
objects have  spectral energy distributions broadly similar  to today's ultra-luminous infrared galaxy 
(ULIRG) population.   In the  local universe the ULIRGs are the most luminous galaxies and emit the bulk of their energy at FIR wavelengths.  The 
FIR 
emission is from dust which is currently thought to be heated by young stars  \citep{lut99}.The ULIRGs are primarily the result of mergers and result 
in 
objects with surface-brightness profiles of 
elliptical galaxies (see \citet{san96} for a review).  Though the dust temperature of the SCUBA sources  is very poorly known, if we assume a 
temperature similar to the local ULIRGS then  these objects are extremely luminous, with  bolometric luminosities of  10$^{12-13}$L$_{\odot}$. 
Radio and CO observations of two SCUBA sources \citep{ivi01} have detected possible extended emission which is in marked difference to  the 
compact nature of local ULIRGs. It is still unclear whether the majority of these objects are powered by star formation or active galactic nuclei 
(AGN).  
Evidence 
is  mounting through X-ray and optical emission line measurements that, although AGN are present in a small fraction of sources, star formation is, 
by 
far,  
the dominant process \citep{ivi00,fab00,bar01}.  Given this, the ULIRGs must be forming stars at unprecedented rates of hundreds to  thousands 
of 
solar 
masses 
per 
year \citep{ivi00,eal99}.   These high star formation rates, together with the contribution that these objects make to the total extragalactic 
background, 
showing this is a cosmologically-significant population \citep{eal00}, makes it hard to avoid the conclusion that these objects are elliptical galaxies 
being 
seen during their initial burst of star formation. \

 Analysis of the spatial clustering of different populations can provide clues to their evolutionary connections. Recent measurements  of the 
clustering of 
two other populations of high-redshift galaxies,  the LBGs and Extremely Red Objects (EROs), \citep{gia98,gia01,dad00,dad01} have yielded 
surprising results.  Studies of the z $<$ 1 universe \citep{lef96,carl00} have found  the clustering strength of galaxies to decrease with 
increasing 
redshift, as  expected in a scenario where structure forms through gravitational instabilities.  However, the LBGs ($z \sim$3) and the EROs ($z 
\gtsim $1) are very strongly  clustered. With hindsight this result is in agreement with the prediction of \citet{kai84}  that the highest peaks in the 
density 
field of the early universe  should be strongly clustered.  At the redshifts of LBGs and EROs the universe was much younger and there had been 
less 
time 
for gravitational collapse. These objects are therefore probably the result of the collapse of the rare high peaks in the density field.  As SCUBA 
sources 
are 
even rarer than the LBGs,  and  based on their star 
formation rates perhaps more massive, they would also be expected to show clustering. \

\citet{pea00} investigated the underlying structure in a submillimeter map of the Hubble Deep Field \citep{hug98}, after removing all 
discrete  sources above 2 mJy, and found no significant clustering of the underlying flux.  However, the HDF area is small and much larger areas 
are needed to investigate the clustering of submillimeter sources. Of the current deep SCUBA surveys there are  two blank-field surveys of 
significant 
size 
and with which a clustering measurement 
may be made: the ``8 mJy survey'' \citep{sco01},  and 
our own, the Canada-UK Deep Submillimeter Survey  (CUDSS).  The ``8 mJy survey''  has detected a clustering signal for $\s8 > $ 8 mJy sources 
over an area of 260 arcmin$^2$. Our survey  covers 100 arcmin$^2$ and reaches a deeper depth of 3 mJy. \

This paper is the sixth of a series of papers on the CUDSS project and contains the complete submillimeter data of our 3\h field.  
The submillimeter survey is now complete and the final 
catalogue contains 50 sources, 27 of which have been detected in the 3\h field. Paper I \citep{eal99} introduces the survey and  initial detections;
paper II 
\citep{lil99} discusses the first optical identifications;  paper III \citep{gea00} discusses the multi-wavelength properties of a 
particularly interesting and bright source, 14-A;  paper IV \citep{eal00} presents the nearly complete 14$^h$ field submillimeter 
sample and  discusses the mid-IR and radio properties of the sources;  paper V \citep{web01} investigates the relationship between SCUBA 
sources 
and 
LBG galaxies 
in 
the CUDSS fields; and papers VII and VIII (Clements et al., in preparation; Webb et al., in preparation) will discuss the optical and 
near-IR properties of the entire sample.    \

This paper is laid out as follows:  \S 2 describes the submillimeter observations, \S 3  discusses the data 
reduction and analysis techniques, \S 4 presents the source catalogue,  \S 5  discusses the radio and ISO data, in \S 6 we discuss individual sources, 
 in 
\S 7 
we present the source counts, 
in \S 8 the clustering analysis is performed and the implications of these results are discussed in \S 9.

\section{Submillimeter Observations}

We observed  60  arcmin$^2$ of the Canada-France Redshift Survey  (CFRS) 3\h field over 25 nights from January, 1998 through July, 2001 
with SCUBA on JCMT \citep{hol99}.  These data are part of the larger Canada-United Kingdom Deep Submillimeter Survey which also includes a 
50 
arcmin$^2$ region in the CFRS 14\h field and two deep 5.4 arcmin$^2$ regions in the CFRS 10\h and 22\h fields.  Some of these  observations 
are 
discussed in earlier papers (please see \S 1 for outline) and will be discussed futher in future papers.  SCUBA is a system of  two  bolometer arrays  
which 
observe   at  850$\mu$m and 450$\mu$m simultaneously.   The beam sizes are roughly 15.0\arcsec  and 7.5\arcsec  at 850$\mu$m and 450$\mu$m 
respectively.  Our data were taken using the ``jiggle mode'' of SCUBA which fully samples the sky plane through 64 offset positions. \

The final image is a mosaic of of 101 overlapping individual jiggle-maps, ($\sim$ 50 minutes per jiggle-map). Each point in the final map contains 
data 
from 
about nine separate jiggle-maps, giving an effective total integration time of about 8 hours.  Because a source will likely fall on different bolometers 
in 
different jiggle-maps, this procedure reduces the chance of spurious sources being produced by noisy bolometers.  This mosaicing procedure also 
produces 
a fairly constant level of noise over a large area of sky. \

We chose a ``chop throw'' of 30 arcsecs so that while chopping off-source an object will still fall on the array, except at the very edges of the 
map.  We chopped in right ascension (RA) which creates a distinct pattern on the map for real objects of a positive source with two negative 
sources (at  half the 
flux) offset by 30\arcsec in RA on either side. In the map analysis this pattern is used to discriminate between real sources and spurious noise 
spikes which, unlike real objects will not be accompanied by two negative sources.   \

The opacity of the atmosphere was determined from ``skydip'' observations which were taken in between each single jiggle-map (approximately 
once 
every hour) except in exceptionally stable weather when skydips were taken every second jiggle-map. Observations to correct for pointing errors 
were  
done with the same frequency  and pointing offsets were consistently less than 2 arcseconds.  The observations were calibrated each night using 
Mars, Uranus, CRL618 or IRC+10216. Although IRC+10216 has long-term variability it has been recalibrated in 1998, and the variablility is much 
less 
than our expected calibration error at 850$\mu$m of approximately 10\%.\

\section{Data Reduction and Source Extraction}

The data reduction procedure is discussed in detail in \citet{eal00}.  We follow the standard SCUBA User Reduction Facility 
(SURF) reduction  procedure \citep{jen97}.  First the ``nod'' is removed by subtracting the  off-position data  from the on-position data.  Next, 
the maps are flat-fielded which removes the effects of sensitivity variations between the bolometers.   We then correct for sky opacity using the 
${\tau}_{850}$  and ${\tau}_{450}$  values determined from the skydips taken before and after each observation.    The median sky value at each 
second is determined and removed from all the bolometers.  This is done because in practice  the chopping and nodding procedure used to remove 
the sky emission does not work perfectly  for two reasons.  First,  the sky may vary faster than the chop frequency.  Second,  this procedure will 
only remove linear gradients in the sky brightness.  Individual measurements from the time-series of each bolometer above 3$\sigma$ are then 
rejected from the data in  iterative steps to reduce the noise.  As our objects are faint, even after 8 hours of integration, we are not in danger of 
removing source flux in  this step.    The data from individual bolometers   are weighted according to their noise  and ``rebinned'' to construct a 
map.  Figure 1 shows our final 850 $\mu$m map, smoothed with a 10\arcsec Gaussian profile.\

The map used for source extraction is produced by convolving this image with a beam template, which includes the negative beams as well as the 
postive beam. We  produced this  from the  observations of our flux calibrators which are close to being point sources.  This suppresses spurious 
sources (noise spikes) in the final map as they are not well matched to the 
negative-positive-negative pattern of the beam.  Real sources, on the other hand, are well matched to the beam template and the flux in the 
negative beams  is 
combined with the positive flux in the final convolved map, thereby increasing the signal-to-noise.  This procedure will also produce a more 
accurate  measurement of the position of the source since  the positional information off all three beams is used.  \

SCUBA acquires data at 450$\mu$m and 850$\mu$m simultaneously and so the 450$\mu$m data, though significantly less useful than the 
850$\mu$m 
data because of the increased sky noise and decreased sky transparancy, comes for free.  The 450$\mu$m data was reduced in the same way as the 
850$\mu$m data and the map was convolved with the beam template. The template-convolved map  is shown in Figure 2.  \ 
 
We generated maps of the noise through Monte Carlo simulations, as described  in paper IV.  For each  jiggle-map data set, we determined the 
standard deviation of the time-stream for each bolometer, and then replaced the time-stream with artificial data with the same standard  
deviation.  We implicitly assumed that the real time-stream has a Gaussian distribution and that the noise on different bolometers was uncorrelated. 
 
The artificial time-stream does not have quite the same distribution as the real time-stream, partly because the reduction procedure gets rid of 
points more than 3$\sigma$ from the mean. To simulate this effect, we applied the reduction procedure to the artificial time-stream, and then 
re-scaled the result so  it again had the same standard deviation as the real time-stream. In 
this way we produced 500 simulated maps and the final noise map (Figure 3) was generated from the standard deviation of each pixel in these 
maps.    \ 

\begin{figure}
\figurenum{1}
\plotone{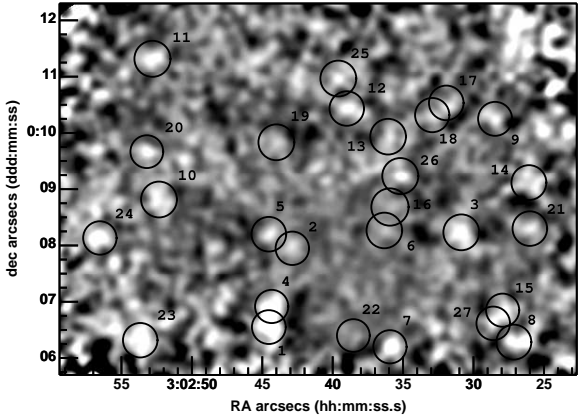}
\caption{The 850 $\mu$m map smoothed with a 10$^{{\prime}{\prime}}$ Gaussian. Object numbers correspond to those in Table 1  and are ordered 
according to  according to signal-to-noise.}
\end{figure}

\begin{figure}
\figurenum{2}
\plotone{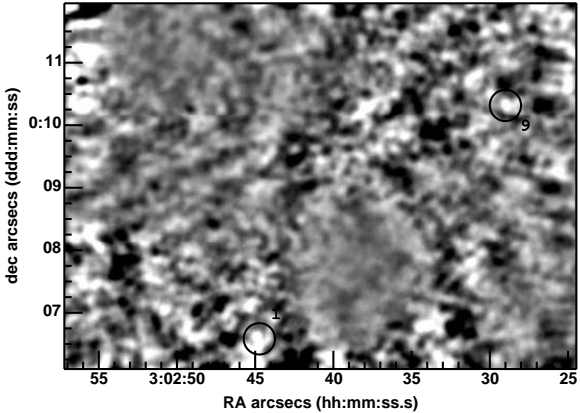}
\caption{The 450 $\mu$m map convolved with the beam template. The atmospheric interference at 450$\mu$m is much worse than at 850$\mu$m 
and the noise pattern is therefore much more obvious. We show the the template-convolved map since it suppresses the noise spikes and the  two 
450$\mu$m detections of 850$\mu$m sources are more obvious to the reader than in a simple Gaussian smoothed map.}
\end{figure}

\begin{figure}
\figurenum{3}
\epsscale{1.0}
\plotone{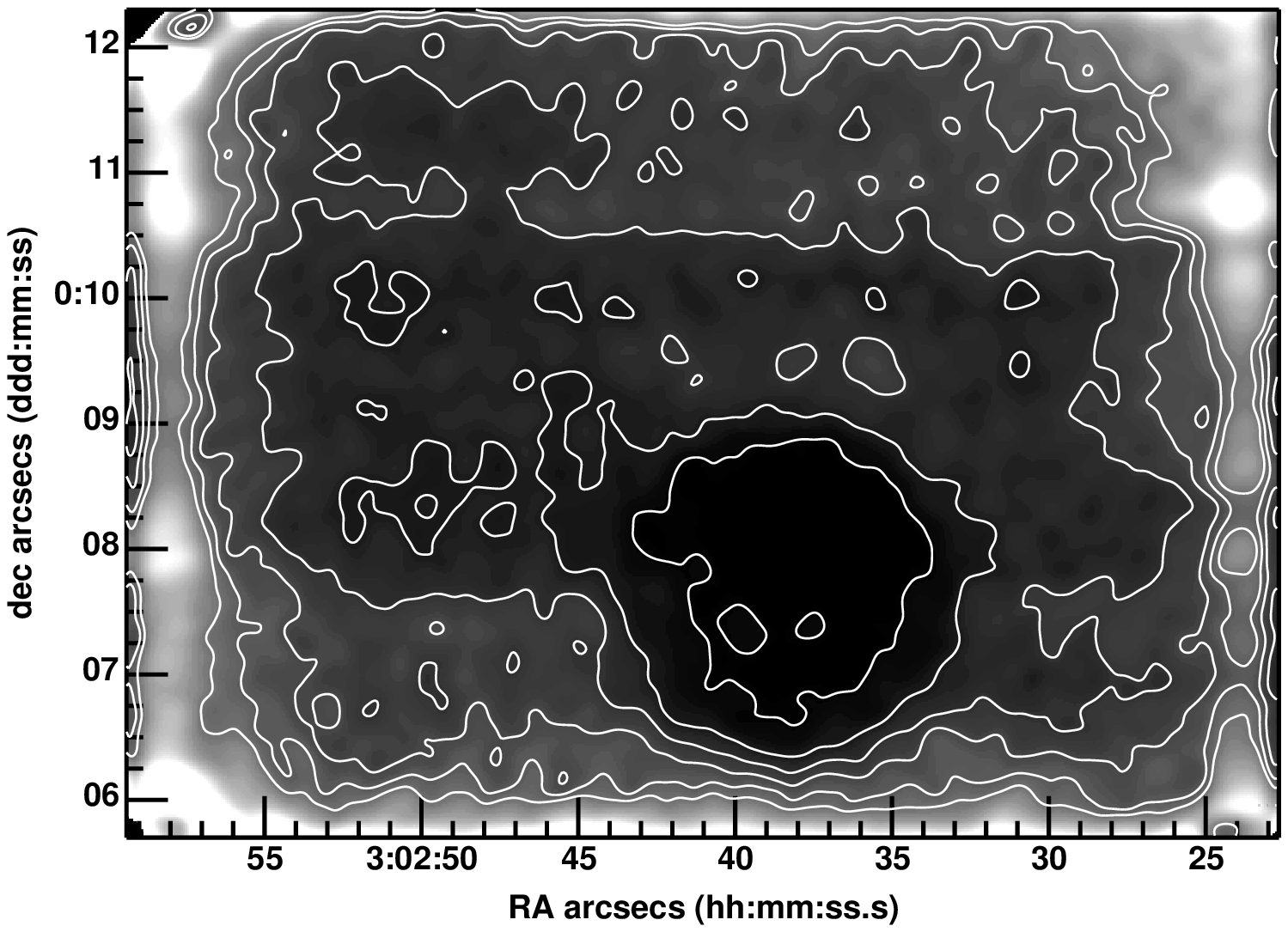}
\caption{The 850 $\mu$m noise map.  The contours begin at 0.75 mJy and increase in 0.25 mJy increments to 2.0 mJy.   The 
non-uniform sensitivity across the field is clearly visible and is due primarily to weather, though the top-most strip has less time coverage. The 
single deep pointing was done in superb weather conditions.}
\end{figure}

Due to varying weather conditions our 3$^h$ map is not as uniformly sampled as our 14$^h$ map (paper III)  and this is clearly visible in Figures 
2 and 3.    Most striking is the single deep pointing (2.7$\arcmin$ across) which was  taken in  early 1998.  The noisy edge effects are also seen 
and a well understood effect of SCUBA's mapping technique.  The center strip of the  map was  observed during excellent conditions while the 
remaining top and bottom strips under more marginal conditions and the top strip has incomplete coverage.  Ignoring the noisy edge  the map may 
 be 
broken into 3 general sections:  

\begin{enumerate}
\item{the central deep pointing: mean 1$\sigma$ noise = 0.77 mJy}
\item{the central strip: mean 1$\sigma$ noise= 1.1 mJy }
\item{the top and bottom strip : mean 1$\sigma$ noise=1.4 mJy}
\end{enumerate}

\section{The Catalogue}

The source extraction was performed on the map which had been convolved with the beam template and then divided by the noise map, giving a 
signal-to-noise map.  The extraction procedure is quite complicated because not only do many of the peaks in the map have low signal-to-noise, 
they 
are 
often merged together.  We used the iterative de-convolution (CLEANing) technique described in Paper IV to compile a source catalogue. \

The data reduction, source extraction, and noise analysis were performed separately and in parallel by the Cardiff and Toronto groups and the final 
source lists compared.  Sources were included in the catalogue if their Cardiff-Toronto average signal-to-noise was 
above  3$\sigma$.  The complete 3\h catalogue is presented in Tables 1 and 2.  There are 27 objects in total which were detected above 3$\sigma$.  
Sources which were less than 3$\sigma$ in one of the maps (but averaged above 
3$\sigma$)  are  noted.  Also listed are offsets in the Cardiff-Toronto  averaged recovered positions.  \

We used the signal-to-noise map to investigate the number of spurious sources.
Since the noise is symmetric around zero 
the  number of  peaks in the inverted map above a given signal-to-noise will be indicative of the number of spurious positive sources.  This   
number  becomes large below 3$\sigma$.  Above the 3$\sigma$ threshold there are 2 negative sources, (both $<$ 3.5$\sigma$) indicating that 
about this  number of 
our positive detections are spurious.  This analysis was done after the CLEANing procedure was complete and full beam profile of the positive 
sources in the map removed.  Therefore the negative sources at $>$ 3$\sigma$ which are not spurious but are associated with real positive peaks 
are not included in this analysis.  The number of spurious sources is approximately the number expected from Gaussian statistics. \   

We define two catalogue lists:  the primary list containing 15 objects above 3.5$\sigma$ which we regard as secure sources and  the secondary list 
containing 12 objects between 
3.5$\sigma$ and 3.0$\sigma$ which are more dubious.   Based on the above reasoning we expect approximately 2 objects in the secondary list to 
be spurious  detections.  \

Also listed in Tables \ref{tbl-1} and \ref{tbl-2} are the 450$\mu$m flux measurements and 3$\sigma$ upper limits for each source, measured at the 
850$\mu$m position on the  450$\mu$m  map.   We searched for 450$\mu$m  flux in two ways.  First, we used the 450$\mu$m map which had 
been 
convolved with the beam  and searched for $>$3$\sigma$ peaks within the error radius of the 850$\mu$m position.  This gave us a list of possible 
450$\mu$m detections.  To test the robustness of these detections we performed aperture photometry on the unconvolved map as in 
\citet{dun01}.  Sources which had consistent flux measurements from both these methods were taken as real detections.  There were two such 
sources corresponding to sources 3.1 and 3.9. This is consistent with the 450$\mu$m number counts presented in \citet{bla99b}. As the  beam of 
the JCMT is 
smaller at 450 $\mu$m (7\arcsec  FWHM) the position of the 
450$\mu$m peak flux is expected to be a better  estimate of the true  position of  the source than at 850$\mu$m and should aid in optical/near-IR 
identifications.  \

\begin{deluxetable}{ccccccc}
\tabletypesize{\scriptsize}
\tablecaption{The 3$^h$ Source List: Objects with S/N $\geq$ 3.5 \label{tbl-1}}
\tablewidth{0pt}
\tablehead{
\colhead{Name} & \colhead{R.A. (2000.0)} & \colhead{Declin. (2000.0)} & \colhead{S/N} &\colhead{S$_{850{\mu}m}$(mJy)} 
&\colhead{S$_{450{\mu}m}$(mJy) } & \colhead{Positional Offset (arcsecs)} \\
\colhead{} & \colhead{} & \colhead{} & \colhead{} & \colhead{}  & \colhead{(3$\sigma$ upper limits)} & \colhead{Cardiff-Toronto ($\prime\prime$)}

}

\startdata
CUDSS 3.1 & 03 02 44.55 & 00 06 34.5 & 7.4 & 10.6$\pm$1.4 & 83$\pm$26 (6.3)\tablenotemark{a} & 1.8 \\
CUDSS 3.2 & 03 02 42.80 & 00 08 1.50 & 6.7 & 4.8$\pm$0.7 & $<$57 & 1.0 \\
CUDSS 3.3 & 03 02 31.15 & 00 08 13.5 & 6.4 & 6.7$\pm$1.0 & $<$63 & 1.8 \\
CUDSS 3.4 & 03 02 44.40 & 00 06 55.0 & 6.2 & 8.0$\pm$1.3 & $<$78 & 0.0 \\
CUDSS 3.5 & 03 02 44.40 & 00 08 11.5 & 5.8 & 4.3$\pm$0.7 & $<$57 & 1.0 \\
CUDSS 3.6 & 03 02 36.10 & 00 08 17.5 & 5.4 & 3.4$\pm$0.6 & $<$33 & 1.0 \\
CUDSS 3.7 & 03 02 35.75 & 00 06 11.0 & 5.3 & 8.2$\pm$1.5 & $<$63 & 1.5 \\
CUDSS 3.8 & 03 02 26.55 & 00 06 19.0 & 5.0 & 7.9$\pm$1.6 & $<$63 & 1.5 \\
CUDSS 3.9 & 03 02 28.90 & 00 10 19.0 & 4.6 & 5.4$\pm$1.2 & 76.9$\pm$25 (0)\tablenotemark{a} & 0.0 \\
CUDSS 3.10 & 03 02 52.50 & 00 08 57.5 & 4.5 & 4.9$\pm$1.1 & $<$57 & 1.0 \\
CUDSS 3.11 & 03 02 52.90 & 00 11 22.0 & 4.0 & 5.0$\pm$1.3 & $<$36 & 1.5 \\
CUDSS 3.12 & 03 02 38.70 & 00 10 26.0 & 4.0 & 4.8$\pm$1.2 & $<$75 & 0.0 \\
CUDSS 3.13 & 03 02 35.80 & 00 09 53.5 & 3.8 & 4.1$\pm$1.1 &  $<$75  & 3.2 \\
CUDSS 3.14 & 03 02 25.78 & 00 09 7.50 & 3.5 & 5.1$\pm$1.5 & $<$63  & 1.8 \\
CUDSS 3.15 & 03 02 27.60 & 00 06 52.5 & 3.5 & 4.4$\pm$1.3 & $<$63 & 1.0 \\
\enddata
\tablenotetext{a}{In brackets is the offset in arcseconds between the 850$\mu$m peak and the 450$\mu$m peak}
\end{deluxetable}

\begin{deluxetable}{ccccccc}
\tabletypesize{\scriptsize}
\tablecaption{The 3$^h$ Source List: Objects with  3.5 $>$ S/N $\geq$ 3.0 \label{tbl-2}}
\tablewidth{0pt}
\tablehead{
\colhead{Name} & \colhead{R.A. (2000.0)} & \colhead{Declin. (2000.0)} & \colhead{S/N} &\colhead{S$_{850{\mu}m}$(mJy)} 
&\colhead{S$_{450{\mu}m}$(mJy)} & \colhead{Positional Offset} \\
\colhead{} & \colhead{} & \colhead{} & \colhead{} & \colhead{}  & \colhead{(3$\sigma$ upper limits)} & \colhead{Cardiff-Toronto ($\prime\prime$)}
}
\startdata

CUDSS 3.16 & 03 02 35.90 & 00 08 45.0 & 3.4 & 2.8$\pm$0.8 & $<$63 & 3.0 \\
CUDSS 3.17\tablenotemark{b} & 03 02 31.65 & 00 10 30.5 & 3.4 & 5.0$\pm$1.5 & $<$75 & 1.8 \\
CUDSS 3.18\tablenotemark{b} & 03 02 33.15 & 00 10 19.5 & 3.3 & 3.9$\pm$1.2 & $<$75  & 1.8 \\
CUDSS 3.19 & 03 02 43.95 & 00 09 52.0 & 3.2 & 3.3$\pm$1.0 & $<$57  & 1.5 \\
CUDSS 3.20 & 03 02 53.30 & 00 09 40.0 & 3.2 & 3.4$\pm$1.1 & $<$57 & 2.0 \\
CUDSS 3.21 & 03 02 25.90 & 00 08 19.0 & 3.1 & 3.8$\pm$1.2 & $<$63 & 0.0 \\
CUDSS 3.22 & 03 02 38.40 & 00 06 19.5 & 3.1 & 3.1$\pm$1.0 & $<$33 & 1.0 \\
CUDSS 3.23\tablenotemark{b} & 03 02 54.00 & 00 06 15.5 & 3.1 & 5.8$\pm$1.9&  $<$78  & 3.2  \\
CUDSS 3.24 & 03 02 56.80 & 00 08 8.00 & 3.0 & 5.1$\pm$1.7 &  $<$60   & 3.0 \\
CUDSS 3.25 & 03 02 38.65 & 00 11 12.0 & 3.0 & 4.1$\pm$1.4 & $<$75 & 2.5  \\
CUDSS 3.26\tablenotemark{b} & 03 02 35.10 & 00 09 12.5 & 3.0 & 3.6$\pm$1.2 &  $<$75 & 1.0  \\
CUDSS 3.27\tablenotemark{b} & 03 02 28.56 & 00 06 37.5 & 3.0 & 4.0$\pm$1.3 & $<$63 & 3.4 \\
\enddata
\tablenotetext{a}{In brackets is the offset in arcseconds between the 850$\mu$m peak and the 450$\mu$m peak, when a clear peak exists}
\tablenotetext{b}{This object is $<$ 3$\sigma$ in either the Cardiff or Toronto map although the average Cardiff-Toronto S/N  is $>$ 3$\sigma$ }
\end{deluxetable}

\section{The Radio and ISO Data}

We obtained a shallow VLA map of the field at 1.4 GHz with the B-array (Yun et al., in preparation) and a preliminary ISO  15$\mu$m map of 
Flores and collaborators (Flores et al., in preparation).
To identify possible associations between the SCUBA sources and ISO and radio objects we used the same positional probability analysis as we use 
for the optical and near-IR data \citep{lil99}.  The probability that an unrelated ISO or radio source will lie within a distance $r$ of a given SCUBA 
position  
can be described by  $P=1-exp(-\pi n r^2)$, where $n$ is the surface density of the ISO or radio sources.  Our Monte-Carlo analysis \citep{eal00} 
implies that the 
true 
position of a SCUBA source will lie within 8\arcsec of the measured position 90-95\% of the time, and we therefore looked for ISO and radio 
sources 
within this distance of each SCUBA source.   Because the surface density 
of ISO and radio sources is  small, we do not expect many chance associations even with such a large search radius. The results are given in Tables 
3 
and 4. 
 \

The source with the largest $P$ value (determined from the radio data) is   CUDSS 3.8, because of the very large offset between the radio 
position and the SCUBA  position. In a sample of 
this size, we would expect about one source to have such a $P$ value,  due purely to chance coincidence.  However, the two other objects with 
radio detections have such small $P$ values that we may regard them a secure identifications.  For the six sources with ISO detections CUDSS 3.8 
is also the least secure, again because of its large offset from the submillimeter position.    Source CUDSS 3.27 is also an insecure identification 
because  we would expect at least one such coincidence in our sample, and it also lacks supporting data, such as a radio detection.   However, given 
the $P$ values of the remaining four ISO identifications we would expect at most one chance coincidence and we therefore regard these 
identifications as secure (in particular the ones also identified with a radio source).

\begin{deluxetable}{cccc}
\tablecaption{Radio Associations}
\tabletypesize{\scriptsize}
\tablewidth{0pt}
\tablehead{
\colhead{CUDSS}  & \colhead{Flux ($\mu$Jy)} & \colhead{offset to SCUBA} & \colhead{P} \\
\colhead{} & \colhead{} & \colhead{position (arcseconds)} & \colhead{} }
\startdata
CUDSS 3.8   & 746 $\pm$ 14 & 9.0 & 0.023 \\
CUDSS 3.10 &  119 $\pm$18 & 1.3  & 0.00061 \\
CUDSS 3.15  & 188 $\pm$ 20 & 2.3 & 0.0019 \\
\enddata
\end{deluxetable}

\begin{deluxetable}{ccccc}
\tablecaption{ISO Associations}
\tabletypesize{\scriptsize}
\tablewidth{0pt}
\tablehead{
\colhead{CUDSS name } & \colhead{ISO name } & \colhead{$S_{15{\mu}m}$ ($\mu$Jy) } & \colhead{offset to SCUBA} & \colhead{P} \\ \colhead{} & 
\colhead{} & \colhead{} & \colhead{position (arcseconds)} & \colhead{} }
\startdata
CUDSS 3.8 & 1003  & 1480 &  9.6 & 0.22 \\
CUDSS 3.10 & 425 & 825 & 1.5 & 0.00087 \\
CUDSS 3.15 & 1040 & 254 & 3.2 & 0.027 \\
CUDSS 3.22  & 770 & 335 & 7.5 & 0.021 \\
CUDSS 3.24 & 382 & 181 & 2.1 & 0.0017 \\
CUDSS 3.27 & 1039 & 174 & 6.7 & 0.11 \\
\enddata
\end{deluxetable}

\section{Notes on Individual Sources}

Below we discuss individual sources with detections at ISO, radio, and 450$\mu$m wavelengths, and sources with Cardiff-Toronto positional 
offsets of $>$2\arcsec.  The optical and near-IR identifications are discussed in detail in Clements et al. (in preparation).

\paragraph{CUDSS 3.1}
This is the brightest object in both  the 3\h and 14\h catalogues.   It is detected at 450 $\mu$m which would suggest it is at low
redshift.  This agrees with a possible optical ID, CFRS 03.0982 (Clements et al., in preparation), which has a spectroscopically determined 
redshift 
of  z=0.1952 \citep{ham95}.    This source is located near the edge  of the field  and highly confused with CUDSS 3.4 as well as a third 
possible 850$\mu$m source at $<$ 3$\sigma$ and so the position is likely to be uncertain.   The  new 450$\mu$m position increases the offset to 
CFRS 03.0982 from 4\arcsec to 7\arcsec, but it remains 
the best optical/near-IR identification.    However, there are two facts which suggest this is not the correct identification. First the probability that 
CFRS 
03.0982 is unrelated to the SCUBA source is quite large (Clements et al. in preparation).  Second, if this 
object does indeed have a redshift of  $z$ = 0.1951 it should certainly be detected at 1.4 GHz, given its 450$\mu$m and 850$\mu$m fluxes,  which 
it 
is 
not.   

\paragraph{CUDSS 3.8}
This is one of three detected in our shallow 1.4 GHz map   and one of six detected with ISO at 15$\mu$m (ISO source 1003).  The optical 
counterpart CFRS 03.0358 has a spectroscopic redshift of $z$=0.0880 \citep{ham95}.   The identification offset of 9\arcsec  is the largest for 
our sample but the radio and ISO  detections and clear merger morphology in the optical image (Clements et al., in preparation) secures the 
identification.  There is no 450$\mu$m detection but this is not entirely surprising since the object lies very near the edge of the smaller 
450$\mu$m map where the noise is high.

\paragraph{CUDSS 3.9}
This object is one of two detected at 450$\mu$m which suggests it is a low-redshift.  However, there is no galaxy visible within 8\arcsec of the 
SCUBA  position in  deep  optical and near-IR images.  

\paragraph{CUDSS 3.10}
This object is identified with ISO source 425  with an offset of 1.5\arcsec. The optical image shows a very bright galaxy (CFRS 03.1299, 
$I_{AB}$=19.4) with clear merger morphology.  It has a spectroscopic redshift of $z$=0.176.

\paragraph{CUDSS 3.13}
The Toronto-Cardiff positions disagree by 3.2\arcsec likely because this source is  extended and may 
therefore be  two sources confused together. 

\paragraph{CUDSS 3.15}
This is identified with ISO source 1040 with an offset of 3.2\arcsec. The optical identification CFRS 03.0346, is a bright ($I_{AB}$=22.1) galaxy 
with  
no 
unusual morphology.    

\paragraph{CUDSS 3.16}
The Toronto-Cardiff positions disagree by 3.0 arcsecs, likely due to its faintness.  

\paragraph{CUDSS 3.22}
This is identified with  ISO source 770, which is  CFRS 03.1029.  

\paragraph{CUDSS 3.23}
 The Toronto-Cardiff position disagree by 3.2\arcsec.  This object is near the edge of the map, where the noise increases rapidly,  and is below 
3$\sigma$ in one of the Toronto-Cardiff 
catalogues.  

\paragraph{CUDSS 3.24}
This source is very close to the edge of the map and has a Toronto-Cardiff positional disagreement of 3.0\arcsec. It is identified with ISO source 
382, which is coincident with a red ($(I-K)_{AB} >$ 3.0) galaxy.

\paragraph{CUDSS 3.27}
 The Toronto-Cardiff positional disagreement for this source is 3.4\arcsec (the largest disagreement in the catalogue) which is not surprising 
since this source is the faintest of a trio of confused sources including CUDSS 3.15 and CUDSS 3.8.   This object is identified with ISO source 
1039, which is CFRS 03.0338.     Note that all three sources in this trio are identified with ISO sources, and one has a radio detection.

\section{The Source Counts}

We present the integral source counts from these data  in Figure 4.  They are in good agreement with the results of other surveys 
\citep{bla99,bor01,sco01}.  To fit the counts we assumed they followed the form  $N(>S) = N_{\circ}S^{-{\alpha}}$ and used the maximum 
likelihood technique outlined in \citet{cra70} to determine $\alpha$.   We find  $\alpha$=3.3$^{+1.4}_{-1.0}$  for the differential counts 
($N(S)\propto S^{-\alpha})$, where the 
errors 
are 95\% 
confidence limits. This is in excellent agreement with  result of  $\alpha$=3.2$^{+0.6}_{-0.7}$  from  \citet{bar99}.   In Paper IV  we investigated 
the effects of  confusion and noise on the source counts and concluded that  the slope is unaffected, though the counts are shifted upwards 
through flux-boosting.  \

In Figure 5 we plot the 3\h and 14\h \citep{eal00} counts.  For the 14\h field we have included four new sources which were not part of our 
catalogue in \citet{eal00}.  Since that paper we have acquired additional 14\h field data and these new sources will be discussed in detail in Webb et 
al. (in preparation).   Although there are more bright objects in 
the 3\h field we see no evidence for a different slope.  Using the same fitting technique we measure $\alpha$=3.7$^{+1.3}_{-1.1}$ for the 
differential counts for the 14\h field.   We note that there was a mistake in the source count analysis of the 14\h field in \citet{eal00}.  However,  the 
conclusion of \citet{eal00} that SCUBA sources brighter than S$_{850{\mu}m} >$ 2mJy contribute  $\sim$20\% to the background at 850$\mu$m is 
unchanged by this new data and the new analysis.

\begin{figure}
\figurenum{4}
\epsscale{1.0}
\plotone{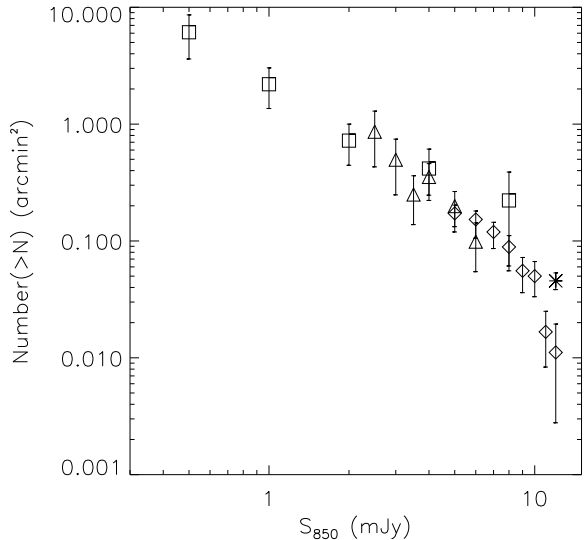}
\caption{The S$_{850{\mu}m}$  source counts determined  by different authors.  The Diamonds correspond to \citep{sco01}, the 
asterisks to \citep{bor01}, the squares to \citep{bla99}, and the triangles to this work using the 3 hour field data. }
\end{figure}

\begin{figure}
\figurenum{5}
\epsscale{1.0}
\plotone{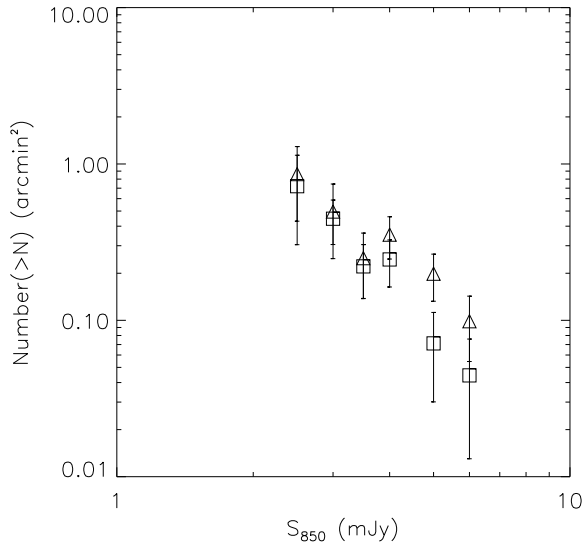}
\caption{The S$_{850{\mu}m}$  source counts  of the two CUDSS fields.  The triangles correspond to the 3-hour field and the squares to 
the 
14-hour field.  Although there are  more bright sources in the 3-hour field the slopes of the counts are consistent with each other.}
\end{figure}

\section{The Angular Distribution  of SCUBA Sources}

We measured the angular correlation function for the SCUBA sources
in the two fields, largely following the procedure given
in \citet{roc99}. To estimate   $\omega(\theta)$ we used the \citet{lan93} formalism:

\smallskip
$$
\omega(\theta) = {DD - 2 DR + RR \over RR} \eqno(1)
$$
\smallskip
\noindent where $DD$ is the number of SCUBA pairs at a given separation, $\theta$, $DR$ is the number of SCUBA-random pairs, and $RR$ 
is 
the number of random-random pairs, all normalized to the same number of objects.  The Landy \& Szalay approach is believed to have good 
statistical properties when a small correlation is  expected.   \

To form $DR$ and $DD$ a sample of 5000 artificial SCUBA sources was generated, carefully taking into account sensitivity variations in the 
images.  To do this we  generated a set of random positions on the assumption of uniform sensitivity across the map.  We then randomly 
assigned fluxes to these position using the best power-law fit to the source counts (\S 7).  Using the noise maps (\S 3) \citep{eal00} we 
determined whether each artificial source would be detected at $>$3$\sigma$ and thereby  modeled the variation in source density across the 
SCUBA  images.   \

The correlation analysis is further complicated by the shape of the  beam.  There will be no sources within $\sim$17 arcsecs  since at 
this 
distance objects will be confused, but the effect of the beam actually extends over separations of $\sim$50 arcseconds in the RA direction (but 
not in Dec.) due to the chop.  Sources can be detected in this area  but the likelihood is decreased. To correct for this  we  placed masks around 
each SCUBA  source corresponding to the shape of the beam.  The number density of random sources around each SCUBA source thus more 
properly reflects  the data.  We calculated ${\omega}({\theta})$ for the entire sample as well  separately for each field. \

A final complication is that of the ``integral constraint''. If $w(\theta)$ is estimated
from an image, the integral 

\smallskip
$$
{ 1 \over \Omega^2} \int \int w_{est}(\theta) d\Omega_1 d\Omega_2
$$
\smallskip

\noindent will necessarily be approximately zero, even though the
same integral of the true correlation function will not be zero
for any realistic image size \citep{gro77}. As in  \citet{roc99}  we assumed the observed angular correlation is given by: 

\smallskip 
$$ 
\omega(\theta) = A(\theta^{-0.8} -C)
$$
\smallskip 

and C is calculated from:

\smallskip
$$
C={\sum \theta^{-0.8}_{ij} \over N_{r_l} N_{r_s}} \eqno(?)
$$

For the 14\h and 3\h fields we measure this to be 0.0106 and 0.0104 respectively. 

Figure 6 shows our estimates of w($\theta$) for both the 3-hour and
the 14-hour field and the combined results where the errors are estimated following \citet{hew82}: 

\smallskip
$$
\delta\omega(\theta)^2 = {1+\omega(\theta) \over N_p}
$$
\smallskip

A visual inspection of the angular distribution of sources in the 14\h field \citep{eal00} immediately suggests clustering, while the 3\h field appears 
 more uniformly distributed, and this is reflected in the correlation function measurement.  We  detect clustering in the 14\h field, but not 
the 3\h and  the clustering detection remains when data from both fields are combined.

Fitting the data for the amplitude we measure $A$=2.4 $\pm$ 4.0 arcsec$^{0.8}$  for the 3 hour field, $A$=6.6 $\pm$ 4.2 arcsec$^{0.8}$ for the 
14 hour field and $A$=4.4 $\pm$ 2.9 arcsec$^{0.8}$ for the combined data.  These two fields are small enough that variance from field to 
field is expected to be important, so this discrepancy is not surprising.   It should be noted though, that the 14\h field, in which we detect marginal 
evidence for clustering, has very uniform noise  properties (due to excellent weather). The clustering signal in the 3\h field may have been washed 
out by the variations in sensitivity across the map which are much more extreme than for the 14\h field.

\begin{figure}
\figurenum{6}
\epsscale{1.0}
\plotone{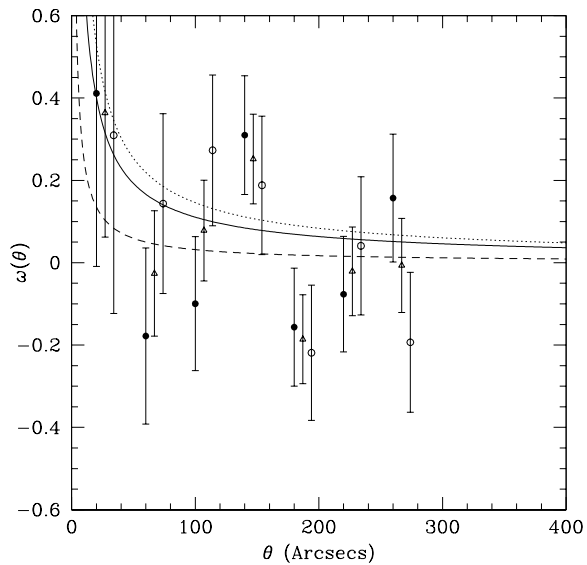}
\caption{The angular correlation function of SCUBA sources. The open circles correspond to the 14-hour field, the filled circles correspond 
to the 3-hour field and the open triangles correspond to the two fields combined. The solid line is the angular correlation function fit to the 
combined data. The dashed line is the angular correlation function of Lyman-break galaxies from \citet{gia98} 
($\omega(\theta)$=2${\theta}^{-0.9}$) and the dotted line  is the angular 
correlation function of EROs \citep{dad00} ($\omega(\theta)$=5.8${\theta}^{-0.8}$).}
\end{figure}

Clustering information could, in principle, be gained from the difference in the number density of the two fields.   The fractional error for the 
integral source counts for both fields is smallest for sources with flux densities greater than 3.5 mJy. We estimated the true surface density of 
objects from  the average surface density of sources in the two fields.  Using this value the difference in the number counts of  the two fields is 
within  the shot noise and is not indicative of clustering.  Still, it does not rule it out  since with such small field sizes 
shot noise is expected to dominate.\

\section{Discussion}

\subsection{The nature and redshift distribution of SCUBA sources}

Determining  the redshift distribution of the SCUBA population has proven difficult for two reasons: (1)  determining the 
optical/near-IR counterpart is not trivial because of the large beam size of the JCMT at 850$\mu$m and (2) the inherent faintness of these 
sources at optical and near-IR wavelengths make spectroscopy difficult.  Consequently, photometric redshift estimates, using a range of 
wavelengths, offer the best alternative.  For the sources presented in this paper we have spectroscopic redshifts for 3/27 and the 450$\mu$m, 
1.4 GHz, and ISO data provide further redshift information. \

The 450$\mu$m to 850$\mu$m flux ratio can be used as a rough redshift estimate, though it is highly dependent on the shape of the spectral 
energy distribution (SED).   In Figure 7 we show the 450$\mu$m to 850$\mu$m flux ratios for these objects as a function of redshift. We include 
the 450$\mu$m detections from the 3\h and 14\h fields and upper-limits from the 3\h field only.   The 
optical counterparts of sources   3.1, 3.8, and 3.10  have spectroscopic redshifts from the Canada-France Redshift Survey (though, we regard 
3.1,  with caution) and for the remaining 
sources (except 3.15 which has a radio detection) we estimated  redshift lower-limits  from their non-detection at 1.4 GHz
\citep{car99,car00,dun00}.  Overlaid are the 
SEDs of three template galaxies.  The solid line corresponds to Arp 220, the archetypal local ULIRG, the dashed line to a reddened starburst 
galaxy  (which has less extinction than a ULIRG) and the dotted line to the more extreme object, IRAS 10214+4724.  To estimate the  reddened 
starburst SED we  used the tabulated values of \citet{sch97} and extended the spectrum to wavelengths larger than 60$\mu$m assuming a dust 
temperature of 48K and a dust-emissivity index of 1.3, a good fit to the starburst galaxy M82.  For IRAS 10214+4724 we assumed a 
temperature of 80K and a dust-emissivity index of 2 \citep{dow92}.  \

 The 450$\mu$m to 850$\mu$m ratio for all three galaxy types drops very rapidly beyond a redshift of $z\sim$1-2 as the observed 450$\mu$m 
flux approaches the peak of the thermal flux.  Thus, a detection at 450$\mu$m is indicative of either a low-redshift or very bright object.  
However, 
the ratio is highly  dependent on temperature and the dust emissivity index  and therefore, in the absence of SED  information loses its power as a 
precise redshift indicator.

In Paper III we found  the  S$_{450{\mu}m}$/S$_{850{\mu}m}$ upper-limits for the 14\h field sources were just consistent with the IRAS 
10214+4724 SED but always consistent with a reddened starburst and Arp 220.  For the sources in  the 3\h field  with estimated redshifts $z>1$
 and   S$_{450{\mu}m}$/S$_{850{\mu}m}$  upper-limits we find a similar result, except for source 3.9 which has an exceptionally high 
S$_{450{\mu}m}$/S$_{850{\mu}m}$ ratio, given its redshift lower-limit.  As in Paper III we measure the mean  450$\mu$m to 850$\mu$m flux ratio 
from the mean 
450$\mu$m flux all sources in the catalogue and find a 3$\sigma$ upper-limit of  S$_{450{\mu}m}$/S$_{850{\mu}m}$ = 2.6.  This is considerably 
lower than for those objects detected at 450$\mu$m (which were included in the estimate) and  is most 
consistent with SED's similar to Arp 220 and the dusty starburst  at redshifts of $z>$ 2.   \

We have detected six of the sources at 15$\mu$m with ISO, including sources 3.8, 3.10, and 3.15 which are  also detected in the radio.  The  
15$\mu$m to 850$\mu$m flux ratio is a much 
stronger function of redshift than the 450$\mu$m to 850$\mu$m ratio (Figure 8) beyond $z \sim$ 0.5 but is also highly SED dependent.  For 
sources  3.8 and 3.10, which have spectroscopic redshifts, $S_{15{\mu}m}/S_{850{\mu}m}$ ratio is lower than found for both the dusty starburst 
and Arp 200. However, the 
remaining four ISO detections appear to have higher ratios than expected for these SEDs if they lie at redshifts greater than $z\sim$3.  At $1.0 < 
z < 
3.0$ 
they are consistent with both Arp 220 and a dusty starburst. \   

The redshift information is listed in Table 5. All redshift lower-limits are derived from the radio/submillimeter flux ratio relation \citep{dun00,car00}.  
The redshift upper-limits are 
based on a detection at 450$\mu$m or 15$\mu$m. We estimated a mean redshift for the entire 3\h sample 
using the routine ASURV \citep{fei85} and find $\overline{z}$=2.0$\pm$0.5. This value, and its uncertainty, was determine using both the 
lower-limit and upper-limit information (separately) and contains only three actual redshift measurements.  We therefore regard it with caution. 
Though it is the lowest mean redshift  measured by the SCUBA blank-field surveys, it is certainly in line with these estimates \citep{dunl01} and 
with our previous estimate from the initial 14\h field data of 2.05$\pm$0.15 \citep{eal00}.Other blank-field surveys have claimed only 10\% of the 
objects lie below $z \sim$ 2 \citep{dunl01}. It is difficult to estimate the number of sources with $z < $ 2 in our sample, however, as we have 3/27 or 
$\sim$ 10\% at $z <$ 1 the number below $z \sim$ 2 is likely much higher given the mean redshift estimate.     \

\begin{deluxetable}{ccc}
\tablecaption{Redshift Estimates for 3\h Field Sources}
\tabletypesize{\scriptsize}
\tablewidth{0pt}
\tablehead{
\colhead{CUDSS name } & \colhead{Redshift } & \colhead{Detection Wavelengths} }
\startdata
CUDSS 3.1 & 0.1952  & 450$\mu$m \\
CUDSS 3.2 & $>$ 1.4 & -- \\
CUDSS 3.3 &$>$ 1.6  & -- \\
CUDSS 3.4 & $>$ 1.7  & -- \\
CUDSS 3.5 & $>$ 1.3 & -- \\
CUDSS 3.6 & $>$  1.7 & -- \\
CUDSS 3.7 & $>$ 1.7 & -- \\
CUDSS 3.8 & 0.0880  & 1.4 GHz, 15$\mu$m \\
CUDSS 3.9 &   1.4 $< z <$ 3.0  & 450$\mu$m  \\
CUDSS 3.10 & 0.176  & 1.4 GHz,15$\mu$m \\
CUDSS 3.11 & $>$ 1.4 & -- \\
CUDSS 3.12 & $>$ 1.4  & -- \\
CUDSS 3.13 & $>$ 1.3  & -- \\
CUDSS 3.14 & $>$ 1.4  & -- \\
CUDSS 3.15 & 1.3 $< z <$ 3.0  &  1.4 GHz, 15$\mu$m \\
CUDSS 3.16 & $>$ 1.1 & -- \\
CUDSS 3.17 & $>$ 1.4  & -- \\
CUDSS 3.18 & $>$ 1.2 & -- \\
CUDSS 3.19 & $>$ 1.2   & -- \\
CUDSS 3.20 & $>$ 1.2  & -- \\
CUDSS 3.21 & $>$ 1.2  & -- \\
CUDSS 3.22 & 1.1 $< z <$ 3.0  & 15$\mu$m \\
CUDSS 3.23 & $>$ 1.5 & --  \\
CUDSS 3.24 & 1.4 $< z <$ 3.0 &  15$\mu$m \\
CUDSS 3.25 & $>$  1.3 & --  \\
CUDSS 3.26 & $>$ 1.2 & -- \\
CUDSS 3.27 & 1.3 $< z <$ 3.0 & 15$\mu$m  \\
\enddata

\end{deluxetable}
\begin{figure}
\figurenum{7} 
\epsscale{1.0}
\plotone{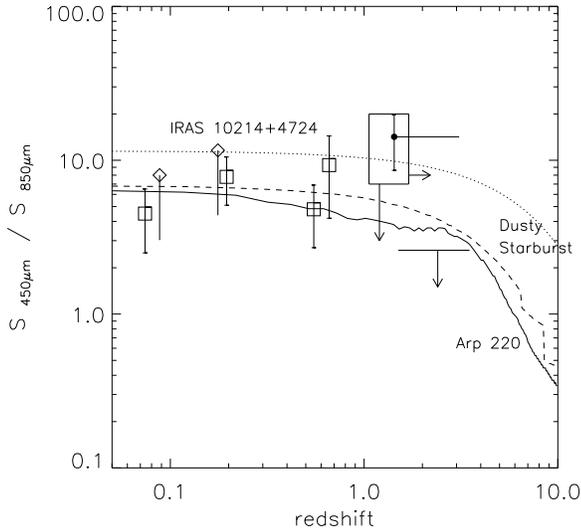}
\caption{The ratio of 450$\mu$m to 850$\mu$m flux as a function of  redshift for the objects detected at 450 $\mu$m in the 3\h, 14\h, and 10\h 
fields \citep{eal00}.  The diamonds represent  
sources 3.8 and 3.10  which have a radio-confirmed identification and a spectroscopic redshift but only an upper limit on the 
S$_{450{\mu}m}$/S$_{450{\mu}m}$ ratio.  The 
square points represent  sources which have  a 450$\mu$m detection and a spectroscopic redshift for the  optical identification.  The 
one 
filled 
circle represents source 3.9  which has lower-limit redshift estimate from its non-detection in our 1.4 GHz radio 
map.  The large square shows the range in  S$_{450{\mu}m}$/S$_{450{\mu}m}$ upper limits and redshift lower-limits (based on the lack of a radio 
detection) for all remaining sources in the 3\h field only. The horizontal line shows the 3$\sigma$ upper limit of the 450$\mu$m to 850$\mu$m ratio, 
derived from the mean 450$\mu$m and 
850$\mu$m flux values for all sources, again for the 3\h field only.}
\end{figure}

\begin{figure}
\figurenum{8}
\epsscale{1.0}
\plotone{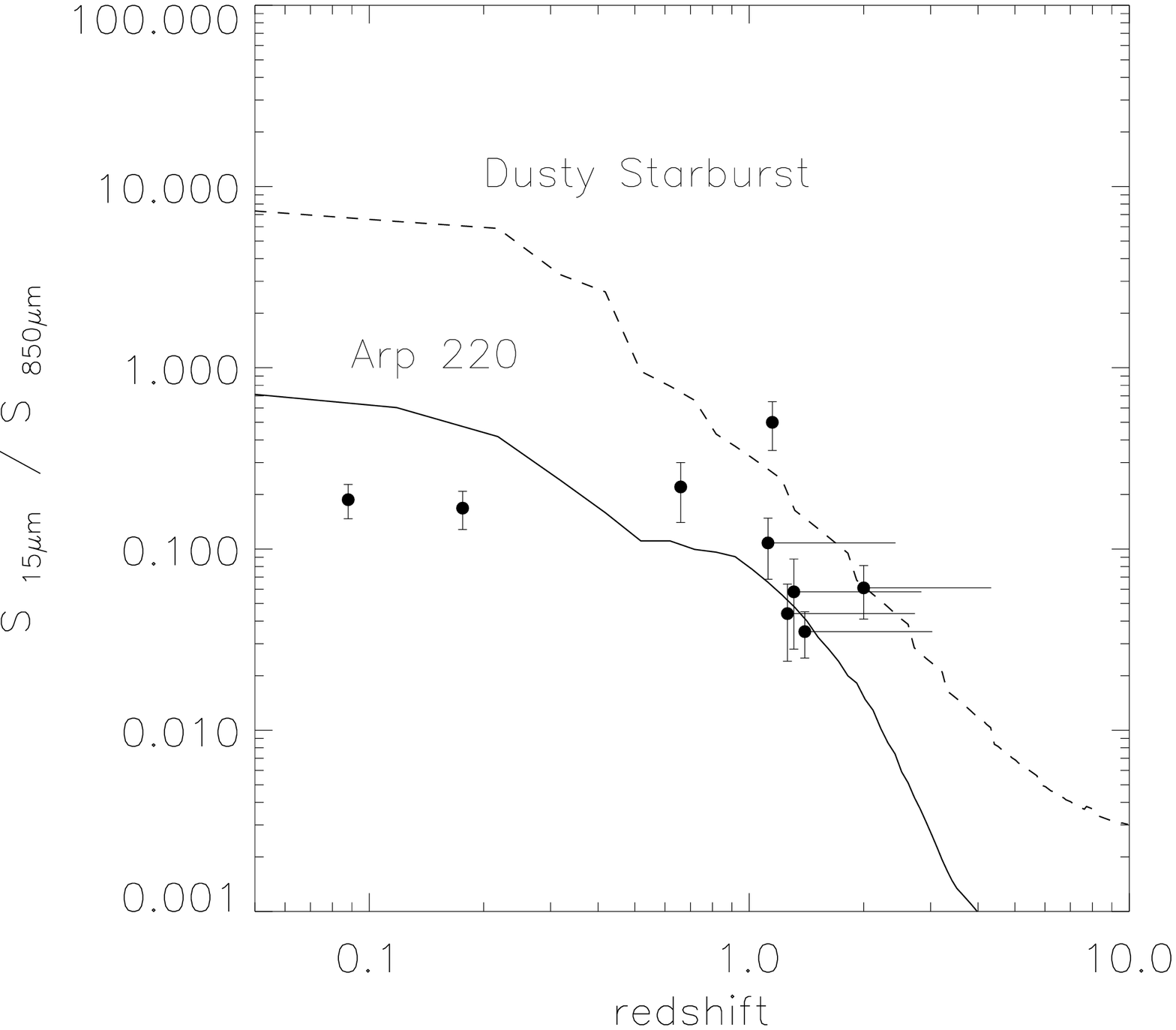}
\caption{The ratio of 15$\mu$m flux to 850$\mu$m flux as a function of  redshift for the sources detected at 15$\mu$m (six in the 3\h field and 3 
in the 14\h field). Overlaid are the SEDs for 
Arp 
200 and a dusty starburst galaxy.  The  objects without spectroscopic redshifts have redshift lower-limits estimated from their non-detection at 
1.4 GHz. }
\end{figure}
\subsection{Clustering of high-redshift dusty galaxies}

Structure formation theory holds that the objects which form from the highest peaks in the density field of the early universe will be strongly 
clustered \citep{kai84}. Indeed, studies of the star forming Lyman-break population, at redshifts $z\sim$3-4 have revealed strong spatial clustering 
\citep{gia98}, implying they formed in the most massive dark halos. At low-redshift clustering strength is strongly correlated with morphology with 
early-type galaxies more clustered than late-types by a factor of $\sim$3 \citep{she01}.   At $z\sim$1,  the Extremely Red Objects are very  
clustered, with $r_o=12 \pm 3 h^{-1}$ Mpc \citep{dad01}.   EROs are  an inhomogeneous mixture of early-type galaxies and dusty starbursts,  
with the 
majority believed to be  massive early-type galaxies  \citep{sti00,mor00}.   Thus, if the objects discovered with SCUBA are progenitors of 
massive spheroidal galaxies we expect them to be clustered as well.   \

Overlaid on Figure 6 are the angular correlation functions derived for Lyman break
galaxies by \citet{gia01}  and for EROs
by \citet{dad00}. For the EROs we have plotted the correlation function for the faintest sample in Table 5 of Daddi et al. Our results are consistent 
with the 
angular clustering of SCUBA sources being as strong
as for either of these populations. \

However, if the angular clustering strengths
of the three populations are the same, their spatial  correlation
functions will have different amplitudes because of their different
redshift distributions. It is difficult to
model this accurately  because, although the redshifts of Lyman-break galaxies are tightly constrained, the redshift distributions of the
SCUBA sources and of the EROs are very uncertain.  It is likely though  that this effect will depress the angular correlation function
of the SCUBA sources relative to the other populations since they are expected to have the widest redshift distribution. \

We can attempt to estimate the spatial clustering strength of the SCUBA sources by assuming a general redshift distribution.
\citet{dunl01} has summarized the current redshift results in the literature and estimates a mean redshift for the SCUBA sources of $z\sim$ 3 with 
$\sim$ 10 \% of the sources at $z <$ 1.   We therefore take the redshift distribution to have a gaussian form, centered at $z=3$ and with a 
standard deviation of ${\Delta}z$=0.8.  Adopting a ${\Omega}_m$=0.3 and ${\Omega}_{\Lambda}$=0.7  cosmology we find, for the combined 
sample, $r_o$= 12.8 $\pm$ 4.5 $\pm$ 3.0 h$^{-1}$ Mpc.  The first error is statistical and is estimated from $\omega(\theta)$ and 
$\Delta\omega(\theta)$.   As ${\omega}_o \sim r_o^\gamma$, and we have fixed $\gamma$ to be 1.8,  a given uncertainty in $r_o$ corresponds to 
a much larger uncertainty  in ${\theta}_o$.  Thus, the largest uncertainty in $r_o$ comes from the uncertainty in the slope of the correlation 
function and is not included in our quoted error.     The second error is systematic and has been estimated by varying the redshift distribution 
parameters:  
$\overline{z}$=2.5-3.5 and $\Delta z$ = 0.6-1.1  (The apparent higher statistical significance of this result is caused by the fact that we have 
assumed 
a 
value for the slope of the correlation function, rather than trying to determine it from the data.  The significance of our result should be taken from 
our 
angular clustering result).   

This value is comparable to the $r_o$=12$\pm$3 h$^{-1}$ Mpc found by \citet{dad01} for EROs but significantly larger than measured for the 
Lyman-break galaxies. \citet{gia01} found values of $r_o$ ranging from 1.0 to 5.0  (with an error of $\sim$1.0 for all) for different UV flux-limited 
Lyman-break samples.  
If SCUBA sources are  indeed more strongly clustered than Lyman-break galaxies, this would suggest  that they  formed in more massive, and 
thus 
rarer, 
dark matter halos  in the early universe.  Indeed, this is the theory put forth by \citet{mag01}.  Using a model first presented in \citet {gra01} they 
suggest 
that  SCUBA sources and Lyman-break galaxies are both  progenitors of QSOs.  In this picture Lyman-break galaxies represent the 
lower-luminosity, 
lower-mass end, and SCUBA sourcees the higher-luminosity, higher-mass end of the same population.  They predict the spatial clustering of 
SCUBA 
sources to be greater than that  of Lyman-break galaxies, and that the clustering strength should increase with submillimeter flux.  At about 
100\arcsec 
they predict 
$\omega(\theta)\sim$0.006 
to 0.02 for $M_{halo}/M_{sph}$=10-100 respectively, for sources with \s8$>$ 1 mJy.  This is much smaller than we could detect with our small 
numbers but also much smaller than our measured value of $\omega(100\arcsec)$=0.11$\pm$0.07 (from our best fit function).\

We do not have  a large enough area to test the  prediction that clustering strength should increase with  submillimeter luminosity, since the 
surface 
density of SCUBA sources drops rapidly with increasing flux.  \citep{sco01} have attempted to measure the clustering of \s8 $>$ 8mJy sources 
over 
a larger area and though they have detected a signal, the small numbers of both our samples make it impossible determine if it is, in fact, larger 
than 
for our lower-flux sample.  

\section{Conclusions}
We have used SCUBA on the JCMT to map 60 square arc-minutes of the CFRS 3\h field.   We have detected 27 sources,  bringing the final 
number 
of objects at \s8$\gtsim$ 3 mJy detected in the CUDSS to 50.  We have found the following results:

\begin{enumerate}
\item{For the differential source counts ($N(S){\propto}S^{-\alpha}$) we measure $\alpha$ = 3.3$^{+1.4}_{-1.0}$ which is in excellent agreement 
with other studies.  Down to 3 
mJy these objects are responsible for $\sim$20\% of the 850$\mu$m background energy}
\item{We have used preliminary ISO 15$\mu$m data, VLA 1.4 GHz observations, and SCUBA 450$\mu$m maps to identify counterparts of the 
850$\mu$m sources.  Using spectroscopy from the CFRS and the radio-to-submillimeter redshift estimator \citep{car99,dun00} we have 
estimated the mean redshift to be 2.0$\pm$0.5 with  10\% of the objects below $z <$ 1.}
\item{We have measured the angular clustering of \s8$>$ 3 mJy sources using the complete CUDSS 3\h and 14\h catalogues.  We find 
$\omega(\theta)=4.4\pm2.9 
\theta^{-0.8}$.  This is as strong as the angular clustering measured for LBGs and EROs, and the spatial clustering will be even stronger 
due to the broad redshift range of SCUBA sources compared to  LBGs and EROs}
\end{enumerate}

{\it Acknowledgments}  
We are grateful to the many members of the staff of the Joint Astronomy Centre who have helped us with this project.  Research by Simon Lilly is 
supported by the National Sciences and Engineering Council of Canada and by the Canadian Institute of Advanced Research. Research by Tracy 
Webb is supported by the National Sciences and Engineering Council of Canada and by the Canadian National Research Council.  Research by 
Stephen Eales, David Clements, Loretta Dunne and Walter Gear is supported by the Particle Physics and Astronomy Research Council.  The JCMT 
is operated by the Joint Astronomy Centre on behalf of the UK Particle Physics and Astronomy Research Council, the Netherlands Organization 
for Scientific Research and the Canadian National Research Council.   We also thank Ray Carlberg for many helpful discussions.   \

\end{document}